\documentstyle[psfig,aa-bib]{l-aa}

\def\pks{PKS\ 2155-304}
\def\bl{BL Lac object}
\def\index{$-2.4$}
\def\sfind{1.3}
\def\z{\phantom{0}}
\def\m{\phantom{--}}
\def\UG{U$_{\mathrm{G}}$}
\def\BG{B$_{\mathrm{G}}$}
\def\VG{V$_{\mathrm{G}}$}
\def\RG{R$_{\mathrm{G}}$}
\def\IC{I$_{\mathrm{C}}$}
\newcounter{count}

\begin{document}
\thesaurus{11.02.2 \pks; 11.16.1}
\title{Very rapid optical variability of \pks}
\author{S.Paltani\inst{1,2,3} \and T.J.-L.Courvoisier\inst{2,3} \and
A.Blecha\inst{2} \and P.Bratschi\inst{2,3}}
\institute{Centre d'Etude Spatiale des Rayonnements, 9, av.\ du
Colonel-Roche, 31029 Toulouse cedex, France
\and
Geneva Observatory, ch. des Maillettes 51, CH-1290 Sauverny, Switzerland
\and
INTEGRAL Science Data Centre, ch.\ d'Ecogia 16, CH-1290 Versoix, Switzerland}
\offprints{St\'ephane Paltani (ISDC). e-mail: {\sf
Stephane.Paltani@obs.unige.ch}}
\date{Received DATE ; Accepted DATE}
\maketitle

\begin{abstract}
We have performed an optical observation campaign on \pks, whose aim
was to determine the variability properties of this object on very
short time scales in several photometric bands. We detected
variability on time scales as short as 15 min.

The Fourier properties of the light curves have been investigated
using structure function analysis. The power spectra are well
described by a power-law with an index \index. It is compatible with
the index found in the X-ray domain. The value of this index shows
that the light curves cannot be generated by a sum of exponential
pulses. Using historical data, we find that the longest time scale of
variability in the optical domain lies between 10 and 40 days.

We find a strong correlation between flux and spectral index, which we
interpret as the signature of an underlying constant component. As a
result we do not find evidence of spectral variation for the active
nucleus in the optical domain.

A lag has been found between the light curves in different optical
bands. The short-wavelength light curves lead the long-wavelength
ones. The amplitude of the lag is about 40 min for a factor 2 in
wavelength.

Our results are compared with predictions from different models. None
of them can explain naturally the set of results obtained with this
campaign, but we bring out some clues for the origin of the
variability.


\end{abstract}

\section{Introduction}
Very soon after their discovery \bl s were noted for their fast
variability. As early as in \citeyear{Racine@70@rapvar},
\citeauthor{Racine@70@rapvar} observed BL Lac with a time resolution
of 15 s and recorded variations of 0.1 mag over a few hours. Since
then, many \bl s and active nuclei of other types are known to vary on
time scales shorter than one day in the optical domain
\cite[e.g.,][]{WagnerWitzel@95@idv}. These objects are usually
classified as \bl s and blazars, and have flat radio spectra. The very
fast variability and the radio spectrum are often interpreted as the
result of synchrotron emission from a Doppler-boosted jet.

\pks\ is the brightest \bl\ in the optical domain, and is consequently
very frequently observed. Variability on time scales of the order of
one day has been observed by \cite{MillerCarini@91@pksrap}. In a
4-night campaign, they obtained a very smooth light curves, which is
well eye-fitted by a unique sine function with a 3-day period. A drop
of $\sim\! 10$\% has been detected in the ultraviolet in a few hours
\cite[]{EdelsonAl@91@rappks}. \cite{TagliaferriAl@91@shortxpks} have
studied EXOSAT observations of \pks, and found that it was dominated
by low-frequency variability, but detected variability on time scales
of the order of 500 s. An important campaign has been performed in
November 1991 with a very broad spectral coverage \cite[][hereafter
1991-I, 1991-II, 1991-III, 1991-IV, respectively
]{UrryAl@93@iuepk,BrinkmannAl@94@xpk,CAl@95@groundpk,EdelsonAl@95@multipk}.
The ultraviolet light curve showed some evidence of characteristic
time scale of $\sim\! 0.7$ days (1991-I). During this period, the
optical and ultraviolet emissions were well correlated, without
evidence of lag (1991-IV). The optical observations have been obtained
from different ground-based telescopes (1991-III), which limited the
accuracy of the observations. Moreover the sampling was not very
dense. As a consequence, the optical variability pattern of \pks\ has
not been fully explored.

In this paper we describe an optical campaign on \pks\ aiming at
determining the properties of the variability of \pks, with an
emphasis on very short time scales. A limited spectral analysis has
also been performed in five-colour photometry to study whether the
hardness of the spectrum is related to the brightness of the
source. We also investigate whether a lag can be detected between the
different light curves. On the basis of our results, we make some
inferences on the emission mechanism of the optical radiation in this
object (and possibly in other \bl s). A preliminary analysis of the
data discussed here have been presented in \cite{PAl@96@pksmiami}.

\section{Observations}
\subsection{The campaign}
\begin{table*}[tb]
\caption{\label{obsdat}General information on the sampling. $n(\Delta
t < 10\mbox{ min})$ and $n(\Delta t < 5\mbox{ min})$ are the number of
consecutive observations separated by less than 10 min and 5 min
respectively. $f_{\mathrm{Norm}}$ is the correction factor applied to
the flux calibration (see Sect.~\protect\ref{flux_cal}). The basic
statistics of the observations are also given. $\sigma$ is the
standard deviation of the light curves in physical units and
$\tilde{\sigma}$ is the standard deviation in percentage of the mean
flux. $\varepsilon_{\mathrm{SF}}$ is the uncertainty on the flux
obtained from the structure functions (see Sect.~\protect\ref{tsa})}
\begin{tabular}{lccccccccc}
\hline
\rule{0pt}{1.2em}Filter&Exposure&Number of&$n(\Delta t < 10\mbox{ min})$
&$n(\Delta t < 5\mbox{ min})$&$f_{\mathrm{Norm}}$&Mean Flux&$\sigma$&
$\tilde{\sigma}$&$\varepsilon_{\mathrm{SF}}$\\
&time (s)&observations&&&&mJy&mJy&\%&mJy\\
\hline
\rule{0pt}{1.2em}\UG&300&26&0&0&1.037&13.3&2.1&15.4&-\\
\BG&180&110&49&0&0.958&15.5&2.3&14.7&0.12\\
\VG&120&262&188&86&0.968&18.9&2.7&14.0&0.13\\
\RG&60&224&144&70&1.052&21.8&2.7&12.2&0.09\\
\IC&80&130&62&8&0.989&24.6&3.1&12.6&0.14\\
\hline
\end{tabular}
\end{table*}

All observations were made with the 70-cm Geneva telescope at the
European Southern Observatory in La Silla, Chile. The telescope is
equipped with a CCD camera using a thick front-illuminated UV-coated
GEC P8603 416x578 device \cite[]{BlechaAl@90@ccd}. Besides the usual
imager mode, the CCD camera can be operated as a multi-channel
photometer. The filters that have been used in this campaign are the
filters U, B and V from the Geneva photometric system (hereafter \UG,
\BG\ and \VG\ to avoid confusion with Johnson's filters; they are
however comparable both in central wavelength and in width), a filter
close to Gunn's R filter (hereafter \RG) and a filter close to
Cousins' I filter (hereafter \IC).

The campaign started on July 26, 1995, and was supposed to last for 3
weeks. However, a bad-weather period prevented us to observe after Aug
9. This resulted in 15 consecutive nights of observation. We observed
\pks\ as often as possible, with a rate strongly dependent on the
filter. Filters \VG\ and \RG\ have been most frequently used. As the
\UG\ observations were difficult to obtain, only one or two of them
were performed each night to increase the spectral coverage. The
exposure times have been estimated {\em a priori} to obtain a 1\%
accuracy on the flux (without taking into account the absolute flux
calibration), which led us to make 2-min exposures in the \VG\
filter. The pointing and tracking limitations of the telescope reduced
the operation period to about 5 hours per night, with a gap of about
45 min when \pks\ was too close to the zenith. Table \ref{obsdat}
summarizes the observations performed during this campaign.

\subsection{Data reduction}
Only the relevant -- user specified -- areas of the CCD are read out,
which enabled us to reduce significantly the delay between two
observations. Images of repeated exposures are stored in a single
structure and reprocessed off-line as follows. The raw flux of the
object is obtained by integration within an aperture of $\sim\! 20$
pixels (8 arc-seconds). The synthetic aperture is centered using a 2D
profile fitting. The sky background is determined individually for
each measurement using an average of the flux outside the integration
zone. The (very rare) cosmic rays are detected and removed using the
comparison with a fitted profile. The flat-field correction is made on
the integrated flux rather than on the raw images (this is equivalent
to the filtering of the flat fields). The data acquisition and image
processing are carried out within the standard Geneva software called
INTER \cite[]{Weber@93@inter}. We perform differential photometry with
a comparison star, which frees us from the problem of atmospheric
extinction (the object and the comparison star are processed in
exactly the same way). The parameters of the comparison star, which
has been well measured in Geneva photometry, are:
$V_{\mathrm{G}}=12.036$, $B_{\mathrm{G}}-V_{\mathrm{G}}=0.104$,
$U_{\mathrm{G}}-V_{\mathrm{G}}=1.263$,
$R_{\mathrm{G}}-V_{\mathrm{G}}=-0.367$ and
$I_{\mathrm{C}}-V_{\mathrm{G}}= -0.752$. Its coordinates are
$\alpha_{2000}= 21\mbox{h~} 59\mbox{m~} 02.35\mbox{s}$,
$\delta_{2000}= -30^\circ~ 10'~ 46.5"$. In order to take advantage of
the differential photometry we centred the field so that both the
object and the comparison star lie in the same frame. Unfortunately,
many observations had to be rejected, because one of the two sources
had partially disappeared from the field of view during the
observation, the angular distance between \pks\ and the comparison
star being close to the size of the field of view.
\begin{figure}[b]
\psfig{file=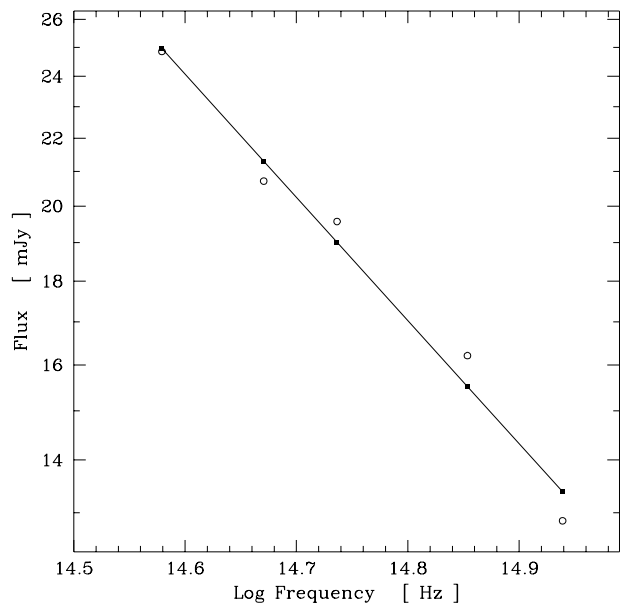,width=8.8cm}
\caption{\label{msp}Mean five-colour spectrum. The empty circles are the
fluxes from the standard calibration, and the black squares are the
fluxes corrected to obtain a power-law spectrum. The slope of the line
is $-0.73$}
\end{figure}

\subsection{Flux calibration\label{flux_cal}}
The absolute flux calibration of the Geneva photometric system
\cite[]{RufenerNicolet@88@calgen} was not intended to be used for
active galactic nuclei. Spectrophotometric observations of \pks\ show
that its spectrum is a featureless power-law (e.g.\ 1991-III). To
obtain absolute flux calibration for \pks, we estimate the most
probable power-law spectrum, given by a least-squares linear
regression, from the mean fluxes obtained with the standard
calibration. The systematic deviations from the power-law spectrum are
then corrected by the appropriate factor (at most 5\%). Fig.~\ref{msp}
shows the mean spectrum with the systematic deviations, and the
correction factors for each filter are given in Table~\ref{obsdat}.

\subsection{Light curves}
\begin{figure}[p]
\psfig{file=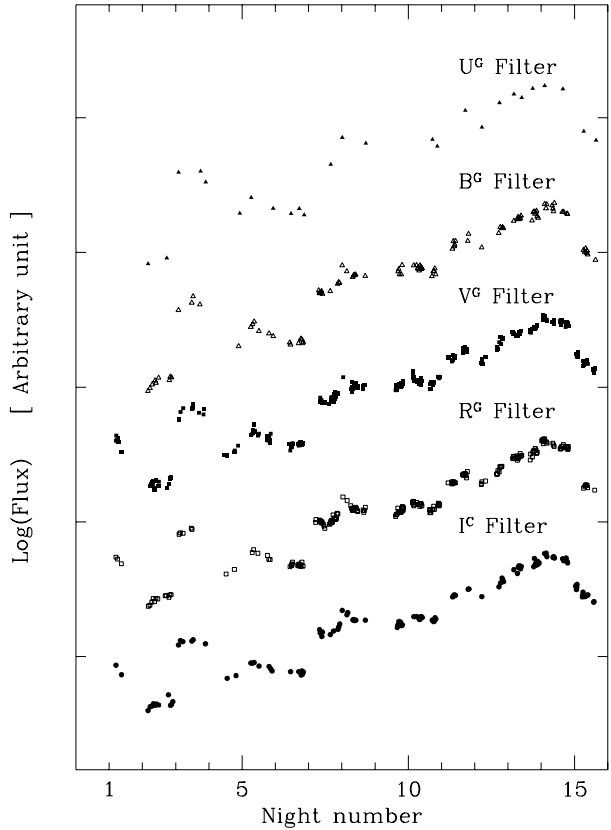,width=8.8cm}
\caption{\label{lc5}Light curves in the 5 filters in arbitrary units. The
light curves are plotted in logarithm of the flux with an arbitrary
normalization. The ticks on the $y$ axis indicate an increase in flux
by a factor 1.5. The $x$ axis is not to scale}
\end{figure}
\begin{figure}[p]
\psfig{file=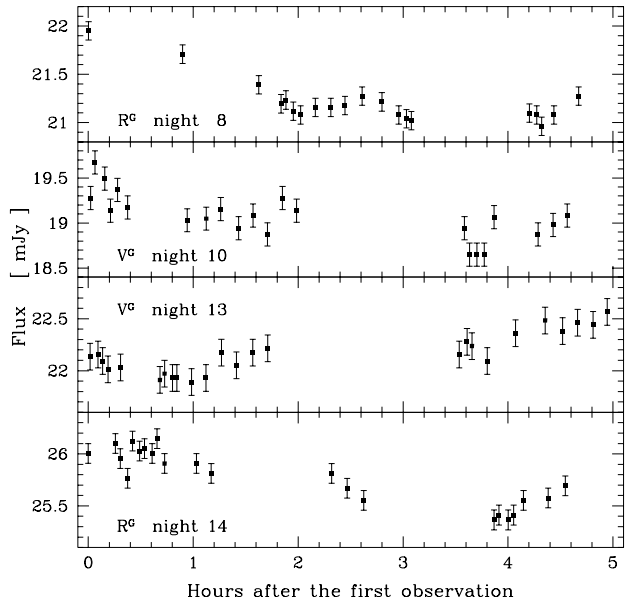,width=8.8cm}
\caption{\label{nig}Examples of nights where very-short-term variability
 is clearly detected. The filter and the night number are given}
\end{figure}

Table~\ref{obsdat} gives the basic statistics of the data discussed
here. Fig.~\ref{lc5} shows the light curves in the five filters in
logarithm of the flux with an arbitrary normalization. All the light
curves look very similar. Fig.~\ref{nig} shows details of 4 nights in
the \VG\ and \RG\ filters. Complicated structures appear during most
of the nights. These structures are characterized by time scales much
smaller than one day. A quantitative description will be given in
Sect.~\ref{tsa}. From visual inspection of the light curves, it
appears that the flux variations are roughly simultaneous, as observed
in 1991-III and 1991-IV. This point will be further analyzed in
Sect.~\ref{delay}.

\section{\label{spa}Spectral index variation}
\begin{figure}[t]
\psfig{file=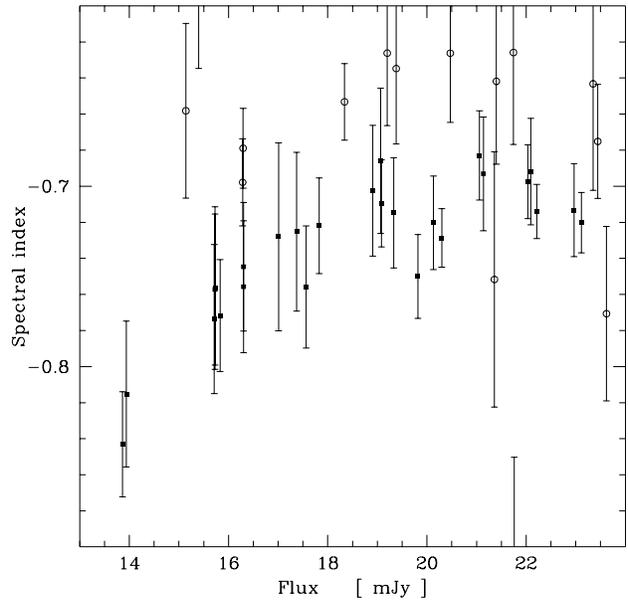,width=8.8cm}
\caption{\label{pla}Spectral index of \pks\ as a function of the flux in
the \VG\ filter. The empty circles are the observations from the 1991
campaign (1991-III)}
\end{figure}

The relative standard deviation of the flux, $\tilde{\sigma}$,
decreases with increasing wavelength (cf.\ Table \ref{obsdat}); the
probability that an uncorrelated population produced a Spearman's
correlation coefficient larger or equal to the measured one ($r=-0.9$)
is about 4\%. This is a first evidence of hardening when \pks\ becomes
brighter.

We have estimated the spectral index $\alpha$ for each observation
with the \UG\ filter. The spectral index as a function of the \VG\
flux is shown on Fig.~\ref{pla}. There is a clear evidence of
hardening when the source gets brighter. A Spearman's test shows that
the probability that no correlation exists between the flux in the
\VG\ band and the spectral index is of the order of $10^{-4}$
($r=0.79$). Note that this kind of correlation is affected by a bias
\cite[]{MassaroTrevese@96@aflbias}. Our result is free from this bias,
because we have followed the recommendations given in the above paper:
the \VG\ band is close to the centre of the \UG-to-\IC\ spectral
region, and the \VG\ flux has not been used in the determination of
the spectral index.

Although the correlation is clear, it disappears completely
(Spearman's $r=-0.04$) if one takes only the observations with a \VG\
flux larger than 18 mJy. The cause of this change of behaviour is
discussed in Sect.~\ref{spectral}.

On Fig.~\ref{pla} we have also plotted the data from the 1991 campaign
for comparison (1991-III). It is obvious that there is a systematic
difference between the two series of spectral indices, the spectral
indices from the 1991 campaign being larger by a constant about
0.08. This difference is probably due to the differences in the
applied absolute flux calibrations, and in particular to the
correction that we have chosen to apply. Although two observations
from the 1991 campaign suggest the existence of the same correlations
for low fluxes, one observation (around 15.4 mJy) is clearly not
compatible with the present data, and another (around 15.1 mJy) is
marginally compatible. We note that at least two other observations
depart from the relationship obtained in the present campaign. This is
discussed in Sect.~\ref{discus}.

\section{\label{tsa}Amplitude and time scales of the variability}

As we have obtained a very dense sampling, we would like to explore
the light curves in the Fourier space. However, we have to avoid the
use of the Fourier transform, since it would be strongly contaminated
by the window function, a function fully determined by the
sampling. Thus we prefer another approach: the first-order structure
function (hereafter we simply refer to the ``structure function'', or
``SF''). The structure function is a tool commonly used in time-series
analysis \cite[e.g.\ ][in clock stability analysis]{Rutman@78@sf}. It
has been introduced in the field of astronomy by
\cite{SimonettiAl@85@sfastro}. The structure function is usually
defined in the case of evenly-sampled discrete time series. The
structure function $S_x(k)$ of an evenly-sampled discrete time series
$(t_i=i\cdot\Delta t;x_i)$, $i=-\infty,\ldots,+\infty$ is given by:
\begin{equation}
S_x(k)= < [x_{i+k}-x_i]^2 >,
\end{equation}
where $<\!f_i\!>$ is the mean of $f_i$ over all $i$. To approximate
this relationship in the case where the time series is given by
$(t_j,x_j)$, $j=1,\ldots,n$ with arbitrary $t_j$, we estimate the SF
in a bin of width $\delta$ for a lag $\tau$ using the relation:
\begin{equation}
S_x(\tau,\delta)= \frac{1}{N(\tau,\delta)} \sum_{(i,j)~|~
\tau-\delta/2<t_j-t_i<\tau+\delta/2} [x_j-x_i]^2,
\end{equation}
where $N(\tau,\delta)$ is the number of couples
$\left[(t_i,x_i);(t_j,x_j)\right]$ that satisfy the relationship
$\tau-\delta/2<t_j-t_i<\tau+\delta/2$.
 
\begin{figure*}[tb]
\psfig{file=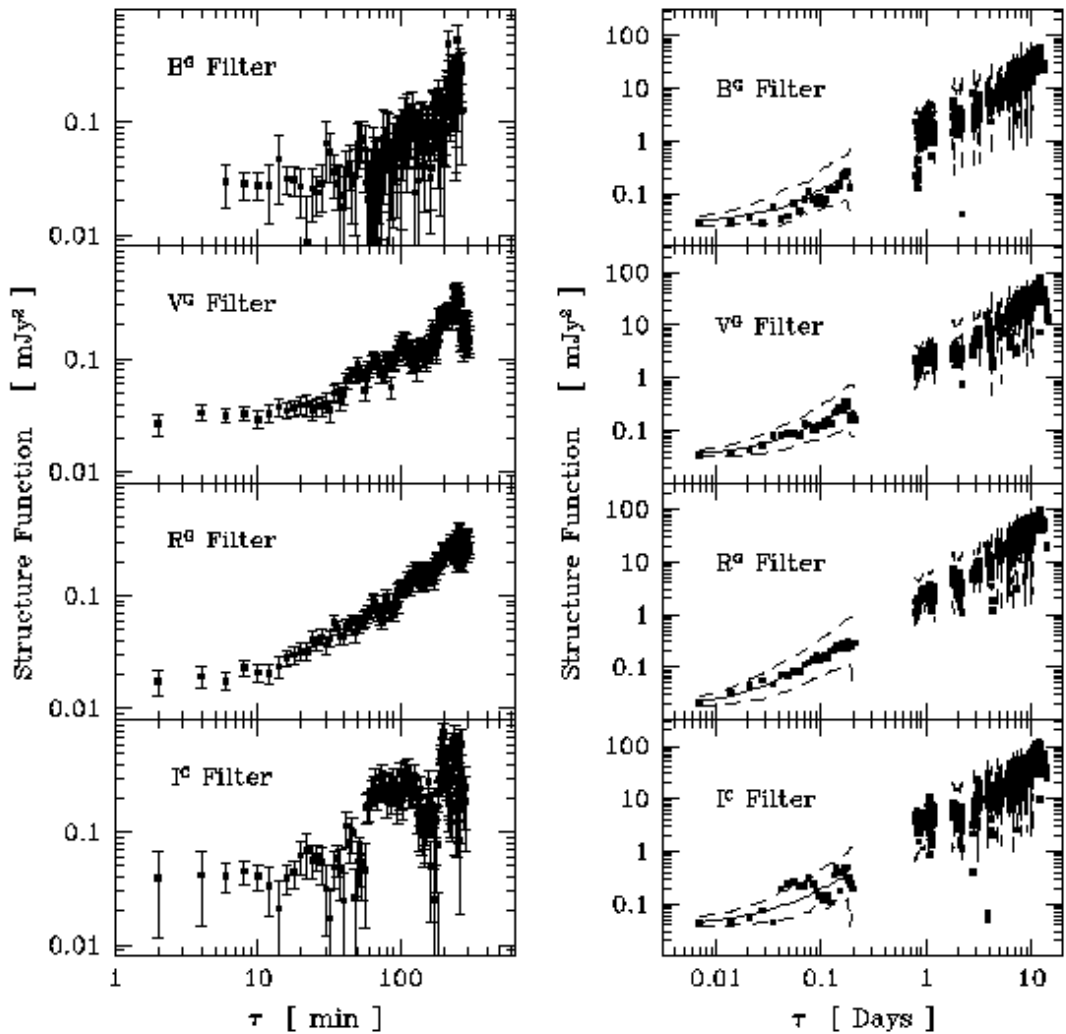,width=16cm}
\caption{{\bf a and b.}\label{sf5} Structure functions of the \BG, \VG,
\RG, and \IC\ light curves: {\bf a} SFs for very short $\tau$, with
$\delta=5$ min. {\bf b} SFs for long $\tau$, with $\delta=10$ min. The
solid lines are the mean SFs of the simulated light curves, and the
dashed lines show the $\pm 1~\sigma$ limits. The SF of the \UG\ light
curve has not been calculated because of the very poor sampling}
\end{figure*}

In the case of the Fourier transform, the effects of a bad sampling
are important, but they can be formulated in an analytical way, and,
in the most fortunate cases, properly corrected using a
deconvolution. For the SF, we do not have such a knowledge about the
perturbations introduced by the sampling. But because of the
particular sampling we have obtained, we expect that they will be much
smaller than in the case of the Fourier transform. Indeed, for some
values of $\tau$, e.g.\ 0.5, 1.5, 2.5,\ldots days, the SF will be
completely unknown, because no observation can be performed 12 hours
after an observation with a single ground-based optical telescope. On
the other hand, for some values of $\tau$, e.g.\ 0.1, 1.0, 2.0,\ldots
days, the number of couples per bin will be very large, and should
lead to a very good determination of the SF. Thus, by preferring the
SF approach, we concentrate the information in some parts of the SF,
instead of spreading it over the complete Fourier domain.

The SFs of the \BG, \VG, \RG, and \IC\ light curves are shown in
Fig.~\ref{sf5}. The SF of the \UG\ light curve has not been
calculated, because of the very small number of observations in this
filter.  Two main features can be observed in the SFs. The first
feature is the presence of a horizontal branch at short $\tau$. This
is due to the fact that the variability of the time series on very
short time scales is dominated by the white noise introduced by the
measurement errors on the fluxes. As the amplitude of a white noise is
independent of the lag between the two observations, the SF of a pure
white noise process is constant, with a value equal to twice the
variance of the white noise. Thus the SF at very short $\tau$ gives an
experimental estimation of the uncertainties on the flux,
$\varepsilon_{\mathrm{SF}}$. These are given in
Table~\ref{obsdat}. The \RG\ light curve, which has the largest
accuracy, shows variability on time scales as short as 15 min ($0.01$
days).

The second feature is a roughly linear increase of the logarithms of
the SFs with $\log\tau$ with a slope about \sfind. An important
property of the SF is that the SF of a time series whose Fourier power
spectrum follows a power-law also follows a power-law. As a first
approximation, we assume that the light curves have power-law shaped
Fourier power spectra. As we cannot correct our SFs from the effect of
the sampling, we use the opposite approach: we simulate 100 continuous
time series whose Fourier power spectra follow a power-law with
different indices. To construct the time series, we simply add
together sine functions with random phases, with the constraint that
the power in each frequency bin decreases as specified. This is a
discrete approximation of a continuous process, but, owing to the very
large number of sine functions added together, we expect that no
serious discrepancy will result from this method. We then project the
simulated light curves onto the sampling obtained for the different
filters, and calculate their SFs. As a result, we obtain the mean SF
{\em for our sampling} of a light curve that is characterized by a
power-law-shaped Fourier power spectrum with a specified index.

The power-spectrum index that matches most closely the observed SFs of
the 4 light curves is \index, the indices between $-2.2$ and $-2.7$
being acceptable (we normalized the simulated light curves so that
their standard deviation are identical to the one of the observed
light curves; therefore the SFs of the simulated light curves match
the one of the real light curve at large $\tau$ for all indices). We
can also remark that the features in the SF of the light curves (e.g.\
the relative drop around $\tau\sim$ 2 days) are not at all significant
when one considers the dispersion in the SFs of the simulated light
curves.

\section{\label{delay}Delay between the light curves}
The existence of a delay between the light curves at different
wavelengths is an important issue, since it has implications on the
emission mechanism. We test this point using two methods.

\subsection{Cross-correlation}
The cross-correlation is the most standard way to obtain a lag between
two time series. We use the interpolated correlation function (ICF)
introduced in \cite{GaskellPeterson@87@gscc}. This method is more
appropriate to our sampling than the discrete correlation function
\cite[DCF, ][]{EdelsonKrolik@88@dcf}, provided that we calculate the
correlations for lags smaller than 0.2 days. The reason is that the
interpolation between two observations made during the same night is a
very good approximation of the actual flux, because of the small
amplitude of variability for very short lags (see
Sect.~\ref{tsa}). Note that \cite{LitchfieldAl@95@simcc} have found
that the ICF method was more efficient than the DCF one when dealing
with simulated flaring light curves.

\begin{figure}[tb]
\psfig{file=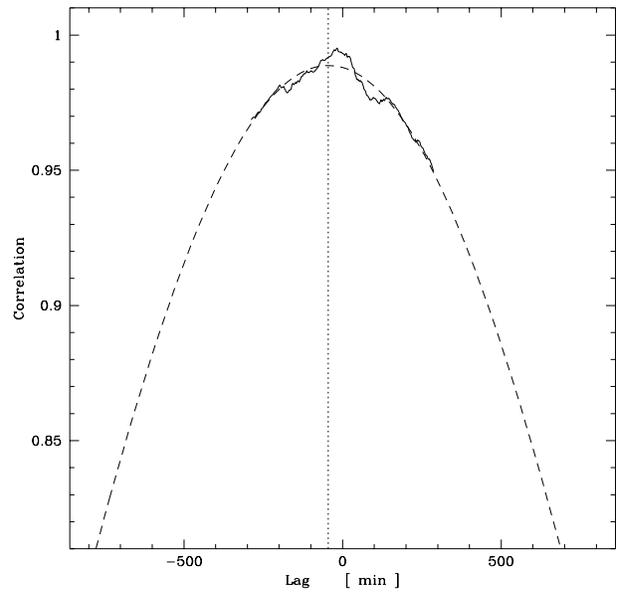,width=8.8cm}
\caption{\label{crb} ICF correlation between the \RG\ and the \BG\ light
curves (solid line). The dashed line is the fit by a Gauss
profile. Its centre is indicated by a dotted line at $\tau=-44.8$ min}
\end{figure}
\begin{figure}[p]
\psfig{file=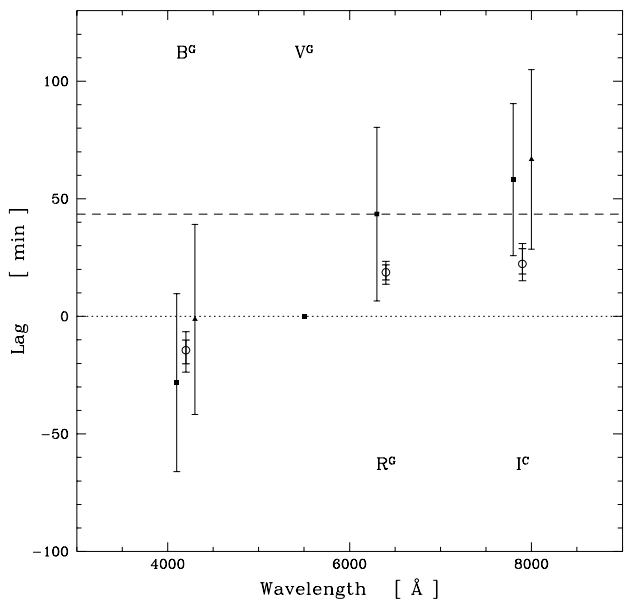,width=8.8cm}
\caption{\label{lag}Lag between the light curves. The black squares are
the lags with the \VG\ light curves (the dotted line is the zero lag
line). The error bars are the dispersions of the lag in the
simulations. The black triangles are the analogous of the black
squares, but with respect to the \RG\ light curves, except that a lag
of $43.5$ min has been added (the lag between the \VG\ and the \RG\
light curves). The dashed line is the zero lag line for the \RG\ light
curve. The empty circles are the results of the global $\chi^2$
minimization method, and give the lag with the \VG\ light curve. The
thick error bars correspond to $\Delta\chi^2=1$, and the thin ones to
$\Delta\chi^2=2.5$}
\end{figure}

The cross-correlations of the different light curves with the \VG\ and
\RG\ light curves obtained with the ICF method all show a very broad
peak that we fitted with a Gauss profile to find its centre
(Table~\ref{cross} and Fig.~\ref{crb}). The peak values are in all
cases very close to 1, which indicates an excellent correlation. It
appears that the shorter wavelengths are leading the longer ones. To
check whether we have really detected a lag, we use the set of 100
simulated light curves with an index \index\ described in
Sect.~\ref{tsa}, that we projected onto the samplings of the real
light curves. About 10\% of the correlations produced a central peak
that could not be fitted by a Gauss profile. This produced aberrant
values for the location of the correlation peak. Instead of analyzing
each of these correlations to obtain a correct value for the lag, we
simply discarded all the values of the lag outside the range $[-0.1;
0.1]$ days. The distribution of the remaining lags is compatible with
a Gauss distribution. The statistics of the location of the peaks are
given in Table~\ref{cross}. As the dispersions of the distributions
are about 8 times smaller than the width of the domain where we
considered the lags as valid, we can conclude that the rejection of
the values out of the domain is justified. The mean lags of the
simulations are different from 0, but significantly smaller than the
dispersion, and are thus neglected.  Fig.~\ref{lag} shows the results
of the cross-correlation analysis.  The significances of the
individual lags are small ($<2\sigma$). However two points indicate
that the lags are most probably real. First, the lag increases
monotonously from the \BG\ to the \IC\ filters. The probability that
this happens by chance is the probability to obtain a 100\%
correlation with a Spearman's test on 4 points, which is about
2\%. Moreover, this has been realized twice independently, since the
same observation can be made for the lags with the \RG\ light
curve. The second point is that the lags between the \BG\ light curve
and the \VG\ and \RG\ light curves are compatible. This is also true
for the lags between the \IC\ light curve and the \VG\ and \RG\ light
curves.

\begin{table}[p]
\caption{\label{cross}Results of the cross-correlations of the observed
and simulated light curves. A negative lag means that the second light
curve is leading. The last two columns refer to the simulated light
curves}
\begin{tabular}{lccc}
\hline
\rule{0pt}{1.2em}Filters\hspace*{3mm}&Observed lag&Mean lag&Dispersion\\
&(min)&(min)&(min)\\
\hline
\rule{0pt}{1.2em}\VG--\BG&--28.2\m&+14.7&37.9\\
\VG--\RG&43.5&--\z9.5&36.9\\
\VG--\IC&58.2&--\z1.4&32.3\\
\RG--\BG&--44.8\m&+12.7&40.4\\
\RG--\IC&23.2&+\z 7.0&38.2\\
\hline
\end{tabular}
\end{table}
\begin{table}[p]
\caption{\label{gcsm}Results of the global $\chi^2$ minimization. A
negative lag means that the second light curve is leading. The
``significance'' is the departure from a zero lag in units of the
standard deviation for the assumptions $\Delta\chi^2=1$ and
$\Delta\chi^2=2.5$}
\begin{tabular}{lccc@{~}ccc}
\hline
\multicolumn{1}{c}{\rule{0pt}{1.2em}Filters}&Lag&Best $\chi^2/$&
\multicolumn{2}{c}{$\Delta\chi^2=$}&\multicolumn{2}{c}{Significance}\\
&(min)&d.o.f.&1&2.5&1&2.5\\
\hline
\rule{0pt}{1.2em}\VG--\BG&--14.4&403.6/371&$\left(^{+4.3}_{-5.8}\right)$&
$\left(^{+7.9}_{-9.4}\right)$&3.3&1.8\\
\VG--\RG&\m18.7&563.9/485&$\left(^{+3.2}_{-3.2}\right)$&
$\left(^{+4.7}_{-5.0}\right)$&5.8&4.0\\
\VG--\IC&\m22.3&486.2/391&$\left(^{+6.5}_{-4.3}\right)$&
$\left(^{+8.6}_{-7.2}\right)$&3.4&2.6\\
\hline
\end{tabular}
\end{table}

\subsection{Global $\chi^2$ minimization\label{chilag}}
The global $\chi^2$ minimization method has been introduced by
\cite{PressAl@92@optlag} to determine the delay between two images of
a gravitationally-lensed quasar. The principle of the method is as
follows: Assuming two unevenly-sampled time series
$(t^{\mathrm{x}}_i,x_i),~ i=1,\ldots,n$ and $(t^{\mathrm{y}}_j,y_j),~
j=1,\ldots,m$, we concatenate them to form a new time series
$(t^{\mathrm{z}}_{\tau_0,i},z_{\tau_0,i}),~ i=1,\ldots,n+m$, one of
the two time series being shifted in time by an arbitrary lag
$\tau_0$. If the covariance properties of the initial time series are
known, one can check whether the concatenated time series has the same
covariance properties. This can be made by calculating:
\begin{equation}
\chi^2=(z_{\tau_0}-\alpha \vec{E})^{T}~ (\vec{C}^{-1})~(z_{\tau_0}-\alpha
\vec{E}),
\end{equation}
where $\vec{C}$ is the total covariance matrix of the process
(including measurement noise), in analogy with the usual $\chi^2$
definition: $\chi^2=\sum_{i=1}^{n}~ (x_i/\sigma_i)^2$. The term
$z-\alpha \vec{E}$ mean that we subtract a constant $\alpha$ to all
components of the $z_{\tau_0}$ vector ($\vec{E}$ is the vector
unity). $\alpha$ is the value that minimizes the $\chi^2$ in the above
equation for an assumed lag, and is given by:
\begin{equation}
\alpha=\frac{\vec{E}^{T}~ (\vec{C}^{-1})}{\vec{E}^{T}~ (\vec{C}^{-1})~
\vec{E}}~ z_{\tau_0}
\end{equation}
It has the purpose of eliminating the bias introduced by the fact the
both the mean and the variance measured in our data are usually
different from their real values. This correction even allows us to
use the method for low-frequency-divergent time series.

The covariance matrix is constructed on the basis of the most complete
structure function drawn as a solid line in Fig.~\ref{sf5}b; it is
important to notice that this covariance matrix is very well
constrained by our data. The details of the construction of this
matrix are to be found in \cite{RybickiPress@92@optrec} and
\cite{PressAl@92@optlag}. A drawback of the method is that it works
only if the $x_i$ and $y_i$ data are two realizations of the same time
series, apart from the existence of a lag, and a possible difference
in mean value. We therefore have to transform the flux in one filter
into the flux that we would have observed in the other filter. We have
observed that the flux in a given filter can be very well approximated
by a constant term plus a linear function of the flux in another
filter, all the correlation coefficients being larger than 0.995.

The delay is then obtained by minimizing $\chi^2(\tau_0)$. The results
are given in Table~\ref{gcsm}, and are plotted on Fig.~\ref{lag}. We
see that the lags are essentially consistent with the ones found with
the ICF method. As it does follow a real $\chi^2$ statistics, we can
in principle estimate the uncertainties on $\tau_0$ by the condition
$\Delta\chi^2=1$. However \cite{PressAl@92@optlag} used the
$\Delta\chi^2=2.5$ after running a set of Monte-Carlo simulations. As
the computation time is huge, we did not try to find the uncertainties
on the lags with our simulated light curves. But, contrarily to the
situation encountered by \cite{PressAl@92@optlag}, the values of the
lag that we have found are in the range where the method should work
well; therefore we believe that the $\Delta\chi^2=2.5$ assumption is
very conservative. The uncertainties are in any case much smaller than
in the previous case. The significances are quite high for all the
lags, even if one uses the conservative assumption.

The probability to obtain such $\chi^2$ values are respectively 12\%,
0.8\%, and 0.07\%. The last two values are very improbable. This may
be due to the fact that this method compares only identical time
series. We used experimental relationships to transform the fluxes in
the filters into ``equivalent \VG\ fluxes''. This can explain easily
the high $\chi^2$ values.

\section{Discussion\label{discus}}

Emission of \bl s is generally supposed to originate from synchrotron
radiation (at least at low frequency). Other mechanisms for
variability have been proposed, like gravitational micro-lensing or
geometrical variations. We are not going to compare detailed models
with the results obtained here, because of their complexity and of the
number of free parameters involved. However our results are strong
enough to constrain the qualitative properties of the models.
\subsection{Time-series properties of the light curves}

We have obtained a very good description of the Fourier power spectra
of the optical light curves of \pks. We can check the compatibility of
our result with the observations from the 1991
campaign. Fig.~\ref{asf} shows the two SFs of the V (considered
identical to \VG) light curve from the 1991 campaign and of our \VG\
light curves.  We see that both SFs are compatible, apart from the
effect of the measurement white noise, which has an amplitude 10 times
lower in our data. This comparison may indicate that the light curves
of \pks\ are stationary time series, or at least that the spectral
properties observed in the present campaign are not completely
peculiar.

Another comparison can be made with the result of
\cite{TagliaferriAl@91@shortxpks}. They found that the Fourier power
spectrum of EXOSAT observations of \pks\ follows a power-law with an
index $-2.5\pm 0.2$, fully compatible with our result. Even after the
removal of the linear trend, the index ($-1.9\pm 0.4$) is still
compatible with our result. This shows that the X-ray and optical
emission are strongly related, and that they have probably the same
origin. However, even in this case, many models predict that the
optical power spectrum will decrease more rapidly than the X-ray one,
the short-time scale variability being attenuated at large wavelength,
mostly because of the longer cooling time (e.g., in synchrotron
radiation). This is not observed in our campaign. These effects, if
they exist, must take place on time scales even shorter than those
investigated here, i.e.\ about 15 min. At least down to this limit,
variability is produced by a wavelength-independent mechanism, for
instance geometrical. This possibility will be rediscussed below.

\subsubsection{Power-spectrum at very high frequencies}
The structure functions of Fig.~\ref{sf5} are all dominated by the
white noise introduced by the measurement uncertainties for $\tau<10$
min. It means that we can only put an upper limit to the minimum
variability time scale in this object. In addition, the structure
function analysis tells us that, if the time between two consecutive
observations is $\tau_0$, the variability in the \VG\ band from the
first observation to the second one is:
\begin{equation}
\sigma_{\tau_0}\simeq \left( \frac{1.4\cdot 10^{-4}}{2} \cdot
\tau_0^{\sfind}\right)^{1/2} \mbox{~mJy},
\end{equation}
where $\tau_0$ is expressed in min. The accuracy should be of the
order 0.02 mJy, i.e.\ 0.1 \% of the flux, to detect variability on
time scales of 5 min. The above formula can help in the planning
of future missions aiming at probing the smallest variability time
scale in \pks. Although very-short-term variability does exist in
\pks, its amplitude is very low. This is due to the steepness of the
power spectrum.

\subsubsection{Power-spectrum at very low frequencies}
\begin{figure}[tb]
\psfig{file=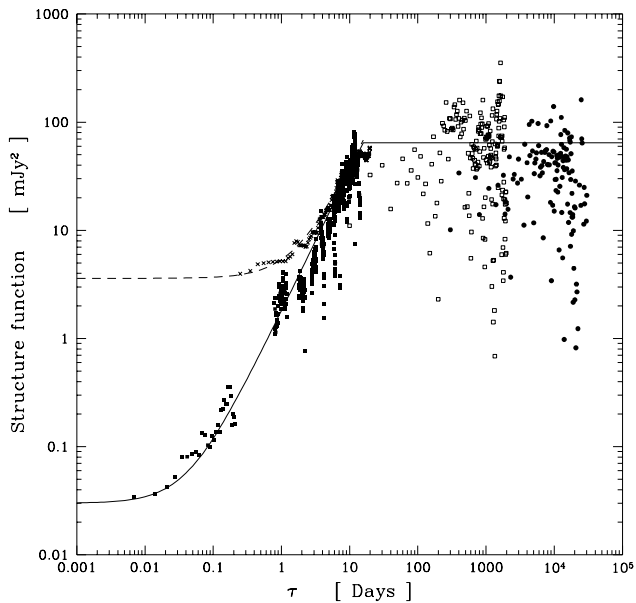,width=8.8cm}
\caption{\label{asf}Structure functions of the light curves in the V band
from our campaign (black squares), the 1991 campaign (crosses), the
Harvard data set (empty squares), and the complete photographic data
set (black circles). The solid line is a power-law with an index
$\sfind$ with a break around 20 days and a contamination with a white
noise with a standard deviation of 0.13 mJy. The dashed line is the
same power-law, but contaminated by a white noise with a standard
deviation of 1.34 mJy. This clearly shows the increase of the accuracy
reached in our campaign}
\end{figure}
As our campaign lasted only 15 days, it is impossible to investigate
the existence of time scales longer than this duration. However there
exists many other observations that can be used to determine the
complete power spectrum. We have used the data from
\cite{GriffithsAl@79@hispk}. The observations are photographic plates
that have been converted into B magnitudes. Their data are presented
in two separate sets: a first set contains the data from the Harvard
photometric collection; it spans about 2\,500 days and contains 66
observations. The second one contains all photographic observations,
but the data have been averaged by year; the set spans about 90 years
and contains 44 annual averages. The B magnitudes have been
transformed into fluxes using a 0-mag flux of $4.27\cdot 10^6$ mJy. To
transform the variations into variations in the V band, we transformed
the B fluxes into \VG\ fluxes using the linear relationship that we
have found between the \BG\ and \VG\ fluxes.

The SFs of the Harvard data set and of the complete photographic data
set are also plotted. The uncertainties on theses SFs are large,
because of the poorness of the sampling, but we see that the SFs of
the two long-term data sets do not follow the increase seen in the
present campaign. As the complete data set is averaged by year, one
must consider its SF as a underestimation of the real SF, because such
an average has great chances of reducing the variability. Taking this
point into account, this SF is compatible with the SF of the Harvard
data. The tentative SF drawn on Fig.~\ref{asf} indicates that the
longest variability time scales in \pks\ lies between 10 and 40 days,
apart perhaps from time scales longer than 100 years.

\subsubsection{Origin of the power spectrum}
The variability of \pks\ is small, and the light curves are very
different from what is observed in, for instance, OJ 287 \cite[e.g.\
][]{Sillanpaa@96@12oj}, which has an amplitude of variability of at
least 4 mag, i.e.\ a factor 40. Actually the variability of OJ 287
appears in sudden bright ``flares'', which could show that the
mechanisms that produce the essential of the variability are very
different from those at work in \pks\ \cite[e.g.\ the binary black
hole model][]{Sillanpaa@96@12oj}. Apart from the distinction between
``flaring'' and ``non-flaring'' objects, we do not know whether the
Fourier properties of \pks\ are typical of \bl s. We have indeed here
the most extensive (in term of spectral coverage) study of the power
spectrum of any active galactic nucleus. The most important point is
that all the data are compatible with a power-law-shaped power
spectrum with an index \index, and and low-frequency cut-off at
frequencies as high as (10-40 days)$^{-1}$.

Many other time series in the physical world have a power-law-shaped
Fourier spectrum \cite[e.g.\ in electronic systems, or, most
surprisingly, in the first Brandenburg Concerto from J.S.\
Bach,][]{VossClarke@75@sound}, but no convincing physical mechanism
has ever been proposed \cite[]{Press@78@flicker}. Several physical
processes have been proposed for X-ray light curves of Seyfert
galaxies. The underlying idea of most of them can be formulated by the
``mechanical model'' of \cite{Halford@68@flicker}. He has shown
analytically that a superimposition of any ``reasonable'' (as defined
in his paper) time-dependent perturbations with different time scales
and amplitude can generate any spectral index in the Fourier
domain. As our index is smaller than $-2$, it means that we can
constrain the properties of the perturbations. It requires indeed that
the Fourier power spectrum of the perturbations decreases with an
index at most \index. Therefore exponential pulses, whose Fourier
power spectrum decreases with an index $-2$, are excluded. A candidate
for the events could be the injection of packets of relativistic
electrons, which cool by emitting synchrotron radiation. Variations in
the parameters of the packets (size, energy distribution of the
electrons, \ldots) could produce the diversity of amplitudes and time
scales required to obtain a power-law-shaped Fourier spectrum (but see
Sect.~\ref{spectral} for the problem of spectral index variation). In
this case the low-frequency cut-off is related to the maximum cooling
time of the packets, if the ``birth times'' of the packets are
completely independent from each other.

Another possibility of producing a power-law-shaped Fourier spectrum
is by summing many periodic functions with different
periods. Following the idea of
\cite{CamenzindKrockenberger@92@lighthouse}, a ``knot'' can have a
helical motion around the magnetic field, which produces periodic
flares, because of the periodic variation of the Doppler amplification
towards the observer. A large enough number of knots with different
gyration radii could fill the power spectrum to produce a
power-law. \cite{CamenzindKrockenberger@92@lighthouse} found that the
typical time scales induced around a $10^8$ M$_\odot$ black hole
should be of the order of 1 day. However it seems plausible that
smaller time scales can be obtained with modifications of the geometry
of the system, like the angle between the jet and the observer, the
velocity of the knot, or the mass of the black hole.

We can further add that it seems rather improbable to us that
gravitational micro-lensing can explain the light curves observed in
our campaign. Even though \cite{KayserAl@89@microlens} have shown that
the light curves generated by a micro-lensing foreground galaxy can be
very complex and have broad Fourier power spectra, their shapes show
successions of sharp flares, which do not appear in the light curves
from this campaign or from the 1991 campaign (1991-I, 1991-II,
1991-III).

Because variability is essentially geometrical (provided that the
cooling times of the packets are significantly longer that the time
scale of the geometrical variability), these last two models predict a
power spectrum mostly independent of the frequency, as observed.
 
\subsection{Spectral behaviour of \pks\label{spectral}}
A correlation between spectral index and flux is clearly seen in our
data. It has however not always been the case in other studies on
\pks. CCD optical observations from \cite{ZhangXie@96@ccdpks} indicate
that there is no correlation between the B-V colour index and the V
magnitude. The complete archive of IUE spectra shows that the
variability increases when the wavelength decreases
\cite[]{PC@94@ulda}, but \cite{Edelson@92@varblazar} found using a
large part of this archive that the flux and the spectral index were
not correlated. On the other hand he found a (not very clear)
correlation in MRK 421. In the 1991 campaign the constant spectral
index hypothesis was favoured in the optical data (1991-III), while a
hardening of the spectrum when the source brightens was the general
behaviour observed in the ultraviolet domain \cite[]{UrryAl@93@iuepk},
but the details were very complex and do not support the clear
correlation found in our data. In other \bl s, the correlation has
been found to be very significant (e.g., OJ 287:
\citeauthor{GearAl@86@iroj} \citeyear{GearAl@86@iroj}; AO 0235+164,
PKS 0735+178, 1308+326: \citeauthor{BrownAl@89@blazariii}
\citeyear{BrownAl@89@blazariii}). On the other hand, in a study of 6
\bl s, \cite{Massaro@95@blaflux} found only two cases of positive
correlations.

Examining Fig.~\ref{pla}, it appears that the spectral index is
compatible with a constant as soon as the flux in the \VG\ band is
larger than 18 mJy. An explanation for the drop of spectral index at
low fluxes could be that the fluxes are contaminated by a constant
emission, e.g.\ from the host galaxy, that is fainter and steeper than
the BL Lac component. \pks\ was rather weak during this
campaign. Indeed the mean V flux obtained here is about 15 \% smaller
than the one obtained during the 1991 campaign. Moreover the V flux
was below 18 mJy during about the half of the campaign, while it was
below this value during about only 20 \% of the 1991 campaign. This,
together with the much better signal-to-noise ratio obtained in our
campaign, could explain why the drop of the spectral index at small
flux has not been observed during the 1991 campaign. A further
argument in favour of the existence of an underlying component comes
from the remark made in Sect.~\ref{chilag}: The best way that we have
found to transform a flux into a \VG\ flux is through a linear
relationship {\em with a constant that increases with the wavelength};
a comparable observation in Seyfert galaxies has been interpreted by
\cite{PW@96@pwrlaw} as the signature of an underlying
component. Therefore we cannot exclude the possibility of a constant
(or weakly varying) spectral index, and it is the interpretation that
we favour.

A few optical observations from the 1991 campaign are in disagreement
with the above interpretation that the variations are achromatic
(apart from the existence of an underlying component). It is also the
case for several of the campaigns on \pks\ cited above. However in the
two optical campaigns, most of the observations are compatible with
the constant spectral index hypothesis (note that a drop of the colour
index can be marginally observed in \cite{ZhangXie@96@ccdpks}). It
suggests that the variability is mostly achromatic (or weakly
chromatic), with some casual significant excursions of the spectral
index from the average value.

Achromatic variability cannot be easily accounted for in terms of
synchrotron radiation, because it would require a very fine tuning
between injection terms and electron losses. On the other hand small
variations of the angle between the jet and the observer, or of the
bulk Lorenz factor can generate achromatic variability. Gravitational
micro-lensing by the stars of a foreground galaxy has already been
invoked to explain the variability of \bl s
\cite[]{SchneiderWeiss@87@agnlens,KayserAl@89@microlens}. However, as
mentioned above, variability is not really achromatic, but rather the
spectral index has little variations, which are uncorrelated with the
flux. This cannot be easily explained either by geometrical variations
alone or by micro-lensing. As the flux variability is rather small, it
may be that spectral index variation is small, if a large number of
packets of electrons with different energy distributions participate
in the optical emission. Moreover a campaign performed in May 1994
that included ASCA observations showed that the variability in the
hard X-ray domain (above the range covered by ROSAT in 1991-II) was
quite different both in amplitude and in time scales from what has
been observed with IUE \cite[]{Urry@96@bvmiami}.

\subsection{Delay between the light curves}
In 1991-IV, \citeauthor{EdelsonAl@95@multipk} already observed a lag
of 2--3 hours between the ROSAT X-ray light curve and the IUE 1400
\AA\ light curve. We confirm the existence of lags between the
different wavelengths, in the sense that the light curves at shorter
wavelengths lead the light curves at longer wavelengths. The delays
between the optical light curves are about 40 min for a factor 2 in
frequency. This value is actually very close (and compatible, owing to
the large uncertainties on these values) to the amplitude of the lag
between the ROSAT X-ray light curve and the IUE 1400 \AA\ light curve
(a factor $\sim 50$ in frequency).

Inhomogeneous jets, i.e.\ where electrons are injected at the base of
the jet and emit most of the synchrotron radiation at frequencies
decreasing with the distance to the base, naturally produce a lag in
the sense observed here (but see Sects~\ref{spectral} and
\ref{summary}). It has however been noted in 1991-IV that the
amplitude of the X-ray--ultraviolet lag is inconsistent with models
where electrons are injected and not reaccelerated, because the life
time of a X-ray-emitting electron is extremely short (of the order of
the second in the model of \cite{GhiselliniAl@85@inssc}). The problem
also exists in the optical domain, where the lag is comparable to the
one between the X-ray and the ultraviolet emissions, and the life time
of the optical-emitting electrons has only increased by a factor of
the order of 10.

This problem can be solved if the lag is due to a bubble of electrons
that becomes progressively optically thin at longer wavelengths
\cite[]{MarscherGear@85@expjet}. We note however that we disagree with
the observation made in 1991-IV that the lag is shorter than the
shortest variability time scales. The 2--3 hours observed in the X-ray
domain are at least 8 times longer that the shortest variability time
scale observed in the \RG\ light curves, and assuredly only the
measurement noise has prevented us to detect even shorter time
scales. This is difficult to explain with this model.

Although stated in the literature (e.g., 1991-IV), we do not consider
plausible that gravitational micro-lensing models can account for the
existence of a lag. The mechanism proposed in 1991-IV should produce a
broad, but symmetrical with respect to $\tau=0$, correlation peak.

Note that the presence of lag between two light curves automatically
implies spectral-index variation. In \pks, the short-time scale
variability is sufficiently small, so that the variation of the
spectral index due to the lag remains undetected.

\section{Summary and conclusion\label{summary}}

\subsection{Results on the variability of \pks}

We have performed several types of analyses to extract the physical
properties of the optical light curves of \pks. The main results are
the followings:
\begin{enumerate}
\item \label{pw}The structure function of the optical variability is
compatible with the one expected for a power-law-shaped Fourier power
spectrum with an index \index.
\item \label{mi}The shortest variability time scale observed is about
15 min, but it is an upper limit, as the variability is dominated by
the measurement white noise below this time scale.
\item \label{ac}The optical spectrum hardens when the source brightens.
We conclude however that this is probably due to an underlying stable
component. As a consequence the variability of the BL Lac component in
\pks\ is considered to be achromatic during this campaign.
\item \label{fc}The fluxes at different wavelengths are very well correlated.
\item \label{la}There is a lag between the different light curves. The
light curves at longer wavelengths are delayed with respect to the
light curves at shorter wavelengths. The amplitude of the delay is
about 40 min for a factor 2 in frequency.
\setcounter{count}{\arabic{enumi}}
\end{enumerate}

To these results, we can add those obtained from the comparison with
other data:
\begin{enumerate}
\setcounter{enumi}{\thecount}
\item \label{ma}The maximum variability time scale is comprised between
10 and 40 days. There is no evidence of variability on time scales
between 40 days and 100 years.
\item \label{xo}The shapes of the power spectra in the optical and X-ray
domain are compatible.
\item \label{vi}The optical spectral index has been observed to vary
slightly without correlation with the flux during other campaigns.
\item \label{lx}The lag between the X-ray and the ultraviolet light
curves is 2-3 hours for a factor 50 in frequency, i.e.\ 20-30 min for
a factor 2 in frequency, which is very close to the lag in the optical
domain.
\end{enumerate}

\subsection{General conclusions}
We have considered the global properties of several models that could
explain the variability of \pks: synchrotron radiation from an
inhomogeneous jet \cite[]{GhiselliniAl@85@inssc}, synchrotron
radiation from an shocked bubble \cite[]{MarscherGear@85@expjet},
variability induced by geometrical variations
\cite[]{CamenzindKrockenberger@92@lighthouse}, and variability induced
by gravitational micro-lensing events
\cite[]{SchneiderWeiss@87@agnlens,KayserAl@89@microlens}. The
following conclusions can be drawn from the above results. They are
sufficiently general to be independent of the details of the models.

\paragraph{Power spectrum origin} Points (\ref{pw}), (\ref{mi}), and
(\ref{ma}) give a very complete knowledge of the variability
properties of the optical light curve. The meaning of the low- and
high-frequency limits of the power spectrum cannot be understood
without a knowledge of the variability mechanism. This is also true
for the shape of the power spectrum. But we can exclude models where
the light curve is generated by the sum of a large number of
exponential pulses, while other shapes of pulses are fully viable. We
can moreover note that some models are less susceptible of generating
the kind of light curves and power spectra observed here, in
particular this seems to be the case for the micro-lensing model.

\paragraph{Power spectrum constancy} Points (\ref{fc}), and (\ref{xo})
indicate that, if the optical power spectrum got through a
high-frequency filter, the low-frequency cut-off of this filter would
be at very high frequencies ($>\!(0.01\mbox{~days})^{-1}$). We prefer
another explanation, namely that the high-frequency variability has a
geometrical origin.

\paragraph{Gravitational micro-lensing models} can be ruled out by Points
(\ref{ac}), (\ref{fc}), and (\ref{la}): Point (\ref{ac}) indicates
that the size of the blue-light emitting region is the same that the
one of the near-infrared-light emitting region, but Point (\ref{la})
shows that these regions cannot be identical, and Point (\ref{fc})
would imply that all lenses cross first the blue-light emitting
region, and about one hour later the near-infrared-light emitting
region with the same amplification parameters. This is, of course,
extremely improbable.

\paragraph{Amplitude of the lag} Points (\ref{la}), and (\ref{lx}) show
that the lag observed between the light curves is much larger than the
one predicted by inhomogeneous jet models without reacceleration. A
model in which a dense bubble subject to synchrotron self-absorption
becomes progressively optically thin can however explain this lag.

\paragraph{Lag -- variability time scale problem} We must remark that
Points (\ref{ac}), (\ref{fc}), and (\ref{la}) are difficult to
reconcile for any model. Models where variability is ``geometrical''
can account for the achromaticity of the optical variability, but it
appears difficult to explain a lag significantly longer than the
minimum variability time scale. This favours an interpretation where
low- and high-frequency variabilities have different origins (e.g.\
synchrotron cooling and geometrical variations respectively).

\paragraph{(Quasi-)achromaticity of the light curves} Synchrotron emission
does not naturally produce achromatic variations, as observed in Point
(\ref{ac}). If synchrotron radiation is the mechanism that produces
the optical emission of \pks, the slight spectral index variations
(Point (\ref{vi})) make us think that the optical emission is produced
by a number of bubbles sufficiently large to ``dilute'' efficiently
the spectral index variations induced by each bubble.\\

The variability picture of \pks\ remains a very difficult puzzle, but
the results given in this paper are very constraining. To our mind,
the most interesting issue raised here is to find a mechanism that can
explain simultaneously important lags and well-correlated variability
on very short time scales. But we keep also in mind that other events
not described here, because not observed during this campaign, exist
in \pks, like the ASCA flare \cite[]{Urry@96@bvmiami}.

\begin{acknowledgements}
SP acknowledges a grant from the Swiss National Science Foundation.
\end{acknowledgements}

\bibliography{biblio}
\bibliographystyle{aa-bib}

\end{document}